\documentclass[a4paper,11pt]{article}
\usepackage{fullpage}

\usepackage{xcolor}
\usepackage[pdfstartview=FitH,
pdfpagemode=UseNone,
colorlinks,
linkcolor={red!60!black},
citecolor={blue!80!black},
urlcolor={blue!90!black}]{hyperref}

\usepackage{graphicx}%
\usepackage{multirow}%
\usepackage{amsmath,amssymb,amsfonts}%
\usepackage{amsthm}%
\usepackage{mathrsfs}%
\usepackage[title]{appendix}%
\usepackage{xcolor}%
\usepackage{textcomp}%
\usepackage{manyfoot}%
\usepackage{booktabs}%
\usepackage{algorithm}%
\usepackage{algorithmicx}%
\usepackage{algpseudocode}%
\usepackage{listings}%
\usepackage{pgfplots}

\usepackage{lmodern}
\usepackage{import}

\everymath=\expandafter{\the\everymath\displaystyle}

\theoremstyle{thmstyleone}%

\theoremstyle{thmstyletwo}%

\theoremstyle{thmstylethree}%

\usepackage{amsmath}
\usepackage{url}
\usepackage{natbib}
\usepackage{booktabs}       %
\usepackage[]{microtype}      %
\usepackage{enumitem}
\usepackage{subcaption}
\usepackage{cleveref}
\usepackage{todonotes}
\usepackage{xcolor}

\setlist[description]{style=nextline}
\usepackage{listings}

\newcommand{\sbit}[0]{\ensuremath{\omega_\mathsf{bit}}}
\newcommand{\faig}[0]{\ensuremath{F_\mathsf{aig}}}
\newcommand{\fbit}[0]{\ensuremath{F_\mathsf{bit}}}
\newcommand{\bit}[0]{\ensuremath{\mathsf{bit}}}
\newcommand{\sol}[0]{\ensuremath{\mathsf{Sol}}}
\newcommand{\tool}[0]{\ensuremath{\mathsf{csb}}}
\newcommand{\csbe}[0]{\ensuremath{\mathsf{csb}}-\ensuremath{\mathsf{exact}}}
\newcommand{\csba}[0]{\ensuremath{\mathsf{csb}}-\ensuremath{\mathsf{approx}}}
\newcommand{\csb}[0]{\tool}
\newcommand{\smap}[0]{\ensuremath{\mathsf{SMTApproxMC}}}
\newcommand{\gnk}[0]{\ensuremath{\mathsf{Ganak}}}
\newcommand{\sstd}[0]{\ensuremath{\mathsf{SharpSAT}\text{-}\mathsf{TD}}}
\newcommand{\arjun}[0]{\ensuremath{\mathsf{Arjun}}}

\newcommand{\tm}[0]{\ensuremath{\mathsf{TechMap}}}
\newcommand{\tse}[0]{\ensuremath{\mathsf{TseitinEnc}}}

\newcommand{\dfour}[0]{\ensuremath{\mathsf{D4}}}
\newcommand{\gpmc}[0]{\ensuremath{\mathsf{GPMC}}}
\newcommand{\vars}[0]{\ensuremath{\mathsf{Vars}}}
\newcommand{\bits}[0]{\ensuremath{\mathsf{bits}}}
\newcommand{\apmc}[0]{\ensuremath{\mathsf{ApproxMC}}}
\newcommand{\qfbv}[0]{$\mathtt{QF\_BV}$}
\newcommand{\tot}[0]{661}
\newcommand{\mxb}[0]{640}
\newcommand{\ecnt}{418}
\newcommand{\ecntp}{643}
\newcommand{\acntp}{659}
\newcommand{\cnts}{111}
\newcommand{\sota}[0]{state-of-the-art}

\newcommand{\cmsg}[0]{\ensuremath{\mathsf{CMSGen}}}
\newcommand{\unig}[0]{\ensuremath{\mathsf{UniGen}}}
\newcommand{\stp}[0]{\ensuremath{\mathsf{STP}}}
\newcommand{\texsf}[1]{\ensuremath{\mathsf{#1}}}

\newcommand{\mparagraph}[1]{\vspace{.5cm}\noindent \textit{#1}}

\usepackage{caption}

\lstdefinelanguage{smtlib}{
  morekeywords=[2]{set-logic, declare-const, assert, check-sat, get-model},
  morekeywords=[3]{proj-var},
  sensitive=true,
  morecomment=[l];,
  morestring=[b]"
}

\usepackage{framed}
\definecolor{shadecolor}{rgb}{0.9,0.9,0.9}

\newcommand{\mpar}{\par\vspace{.5cm}\noindent}

\begin{document}
\title{\bf CSB: A Counting and Sampling tool for Bit-vectors\thanks{This is the authors' version of the article published in Acta Informatica \cite{SM26}. The corresponding tool {\csb} is available at \protect\url{https://github.com/meelgroup/csb}.}}
\author{
\textbf{Arijit Shaw}\\
Chennai Mathematical Institute\\
IAI, TCG-CREST, Kolkata
\and
\textbf{Kuldeep S. Meel}\\
Georgia Institute of Technology\\
University of Toronto
}

\date{}

\maketitle

\begin{abstract}
  {}
  Satisfiability Modulo Theory (SMT) solvers have significantly advanced automated reasoning due to their effectiveness in solving problems across various fields. With the advancement in SMT solvers, there is growing interest in exploring capabilities beyond mere satisfiability, similar to the progression observed in Boolean satisfiability solvers that expanded into counting and sampling. In this study, we investigate the following question:  \textit{Can we rely on modern CNF model counters and CNF samplers to extend modern SMT solvers to handle the problems of counting and sampling over bit-vectors?}

  The main contribution of this work is the development of an efficient and user-friendly tool, {\csb}, that solves a bunch of problems around model counting and sampling on the theory of bit-vectors, namely exact and approximate projected and non-projected model counting, along with the almost-uniform and uniform-like sampling. In the case of exact counting, projected counting, and uniform sampling, {\csb} is the first tool to solve the problem\,---\,although all these problems have a lot of applications.

  Our tool {\tool} converts the bit-vector formula into a CNF formula using bit-blasting techniques before applying CNF model counters or samplers to perform counting or sampling. It keeps track of the variable mapping between the bitvector and CNF formula and passes that information to the CNF counter. We built our tool on top of SMT solver {\stp} by integrating approximate model counter {\apmc}, exact model counter {\gnk},  almost-uniform sampler {\unig}, and uniform-like sampler {\cmsg} in it. Our experiments demonstrate significant performance improvements over existing methods.

\end{abstract}

\section{Introduction}

The paradigm of Satisfiability Modulo Theory (SMT) solving has been central to advances in hardware and software verification over the past two decades. Over time, the community has developed several scalable state-of-the-art SMT solvers~\cite{cvc5,BSST21,boolector,mathsat5,bitwuzla}. For many problem instances, satisfiability alone does not suffice, and one is often interested in computations over the solution set. Two such problems are computing an estimate of the cardinality of the solution set and uniformly sampling a solution from the entire solution set. As a starting point, we focus on the case when the underlying formula is expressed in \emph{{\qfbv}, or quantifier-free bit-vector arithmetic} (referred to as \emph{bit-vectors} hereafter). Our choice of {\qfbv} stems from its being one of the first theories to be investigated in the context of SMT solving, as well as recent empirical studies showing the importance of counting problems over bit-vector formulas in domains such as cryptography~\cite{BZG20} and software verification~\cite{GFS21, TW21}. In many of the problems like ~\cite{TW21}, the underlying problem becomes a \textit{projected model counting} problem, where we want the count projected on a subset of variables.

The problem of counting and sampling over bit-vectors can be addressed through two methods: (i) reasoning directly over bit-vectors, or (ii) reducing the problem to a Boolean formula in conjunctive normal form (CNF). While the former approach has been studied in recent years using lifting techniques for word-level constraints~\cite{CDM15, CMMV16, DBS18, DBS19}, the latter has not been thoroughly assessed. In this paper, we focus on whether a modern SMT solver can be extended with CNF-based model counters and CNF samplers to efficiently address the problems of counting and sampling over bit-vectors.
Our investigation into the design of a counting tool for bit-vector formulas relies on bit-blasting, followed by the use of CNF-based samplers and counters, leveraging the latter's scalability.

Our tool \csb\footnote{Open source tool available at: \url{https://github.com/meelgroup/csb/}} addresses six key challenges in the theory of bitvectors: exact and approximate model counting, exact and approximate projected model counting, uniform‐like sampling, and almost‐uniform sampling. Prior to \csb, no existing tool supported this full spectrum of functionality beyond approximate model counting. To evaluate scalability, we assembled a suite of \tot\ benchmarks drawn from diverse application domains—such as cryptography and software verification—each encoded as a bit‐vector model counting instance.

The primary contribution of this paper is \tool, a high‐performance bit‐vector model counter that extends the SMT solver \stp\ by integrating off‐the‐shelf CNF counters. On our \tot\ benchmarks, \tool\ computes exact counts for \mxb\ instances, outperforming the state‑of‑the‑art counter \smap, which solves only \cnts\ cases. We further enhance \csb\ to support both exact and projected model counting, and incorporate CNF‑based sampling engines to enable efficient uniform‑like and almost‑uniform sampling. Experimental results show that \tool\ generates 500 uniform‑like samples in an average of 6.6 s and produces 500 almost‑uniform samples in 283.7 s.

\mparagraph{Organization.}
The rest of the paper is organized as follows: We introduce the preliminaries and related work in \Cref{sec:background}. In \Cref{sec:framework}, we present an overview of our framework, {\tool}. We describe our experimental methodology and results in \Cref{sec:results}. Finally, we conclude in \Cref{sec:concl}.

\section{Background} \label{sec:background}

\textit{SMT and bitvectors.}
A \emph{bit-vector} is a vector of bits of a given fixed width. Let $X$ be the set of bit-vector variables, and let $F$ be a formula in the theory of \emph{quantifier-free bit-vectors}. Suppose $F$ has variables $x_1, x_2, \dots, x_n$, each associated with a fixed width $w_1, w_2, \dots, w_n$. Consequently, each variable $x_i$ ranges over a finite domain of cardinality $2^{w_i}$. A \emph{model} (or \emph{solution}) of $F$ is an assignment of bit-vector constants to the variables in $X$ such that $F$ evaluates to true. The set of all models of $F$ is denoted by $\sol(F)$. Throughout this paper, we abuse notation by using $F$ to denote both Boolean and bit‑vector formulas interchangeably.

\newcommand{\solproj}[2]{\ensuremath{\mathsf{Sol}(#1)_{\downarrow{#2}}}}
\newcommand{\solprojf}[1]{\solproj{F}{#1}}

\mpar
\textit{Model Counting.}
 The model counting problem is determining $|\sol(F)|$, where $|S|$ denotes the cardinality of a set $S$.
\textit{An exact model counter} takes in formula $F$, and  returns $|\sol(F)|$. \textit{An  approximate model counter} takes in a formula $F$, tolerance parameter $\varepsilon$, confidence parameter $\delta$ and returns $c$ such that
$\Pr\left[\frac{|\sol(F)|}{1+\varepsilon} \leq c \leq (1+\varepsilon) |\sol(F)| \right] \geq 1-\delta$, where $\Pr[E]$ denotes the probability of event $E$.

\mpar
\textit{Projected Model Counting.}
Let $F$ be a formula over variables $\vars(F)$, and let $\alpha \in \{0,1\}^{\vars(F)}$ be an assignment.  For any projection set $S \subseteq \vars(F)$, we define a projected assignment $\alpha$ on $S$
$\alpha_{\downarrow S} = (\,\alpha(x)\,)_{x \in S} \in \{0,1\}^S$
i.e.\ the projection of the full assignment $\alpha$ onto the coordinates indexed by $S$.
Let $\solproj{F}{S}$ denote the set of projected assignments satisfying the given formula $F$ and a projection set $S$. The problem of \textit{projected model counting} is to compute $|\solproj{F}{S}|$.
\textit{An exact projected model counter} takes in formula $F$, and  returns $|\solproj{F}{S}|$. An  \emph{approximate projected model counter} takes in a formula $F$, projection set $S$, parameters $\varepsilon$, and $\delta$, and returns $c$ such that $\Pr\left[\frac{|\solproj{F}{S}|}{1+\varepsilon} \leq c \leq (1+\varepsilon) |\solproj{F}{S}| \right] \geq 1-\delta$.
To differentiate between model counting and projected model counting, we use the term \textit{non-projected model counting} to refer to model counting without projection.

\mpar
\textit{Sampling.} An \emph{almost uniform sampler} $G$ takes a tolerance parameter $\varepsilon$ along with $F$ , and  guarantees $ \forall y \in \sol(F), \frac{1}{ (1+\varepsilon) |\sol(F)|} \leq \Pr [G(F,\varepsilon) = y] \leq \frac{(1+\varepsilon)} {|\sol(F)|}$. An \textit{uniform-like sampler} also takes in a Boolean formula $F$ and returns $\sigma \in \sol(F)$, and is designed to behave like a uniform sampler, but without theoretical guarantees. Its design is inspired by the distribution‐testing tool Barbarik~\cite{CM19,GSCM21}. Empirical evaluation on a large benchmark suite shows that distribution-testing tools cannot distinguish its output from that of an almost-uniform sampler, indicating that the uniform-like sampler produces samples from a distribution closely approximating the ideal almost-uniform distribution.

\mpar
\textit{Independent Support.} For a given assignment $\sigma$ over $X$ and a subset of variables $S \subseteq X$, let $\sigma_{\downarrow S}$ represent the assignment of variables restricted to $S$. Given a Boolean formula $F$ over the set of variables $X$ and a projection set $S \subseteq X$, a subset of variables $\mathcal{I}$ such that $\mathcal{I} \subseteq S$ is called independent support of $S$ if $\forall \sigma_1, \sigma_2 \in \sol(F), \sigma_{1\downarrow\mathcal{I}} = \sigma_{2\downarrow\mathcal{I}} \implies \sigma_{1\downarrow S} = \sigma_{2\downarrow S}$.
Several preprocessing techniques for model counting have been proposed, which compute a small independent support for the input formula and simplify the formula based on that support~\cite{LLM16,SM19}.

\paragraph{Related Work.}
The success of propositional model counters, especially approximate ones, led to the adaptation of these techniques for word-level constraints. ~\cite{CDM15} extended hashing-based model counting from the approximate CNF model counter {\apmc}~\cite{CMV13} to word-level benchmarks using bit-blasting. ~\cite{CMMV16} developed {\smap}, which lifts hash functions to handle word-level constraints. Kim et al.~\cite{KM18,KM20} introduced a system for statistical estimation to continuously refine the probabilistic estimate of model counts, although it lacks $(\varepsilon,\delta)$-guarantees~\cite{KM18}.

In the sampling domain, the need for uniform samplers and those achieving high coverage is well-established. There have been considerable efforts to sample from SMT formulas with high coverage~\cite{DBS18, DBS19, PRI23}, yet uniform sampling from these formulas remains largely unexplored, representing a significant challenge. The technique of lifting hash functions, as used in \cite{CMMV16}, is ineffective for uniform sampling due to the need for 3-wise independence, whereas the lifted hash functions in \cite{CMV16} only ensure 2-wise independence.

In the context of CNF counting, there has been a significant improvement in the last two decades. Different model counters have their respective strengths and weaknesses, as seen in recent studies~\cite{SM24} and the model counting competition~\cite{FHH21}.

\section{Approach}
\label{sec:framework}

We develop {\tool}, a model counting and sampling tool for the quantifier-free fragment of the theory of bit-vectors. To achieve this, {\tool} first bit-blasts the formula to a Boolean CNF formula, then applies an off-the-shelf CNF model counter or CNF sampler to solve the problems of model counting and sampling, respectively. By reducing the problem to CNF, our tool {\tool} leverages ongoing improvements in propositional model counting and sampling.

\begin{figure*}[htb]
  \centering
  \includegraphics[width=0.95\textwidth]{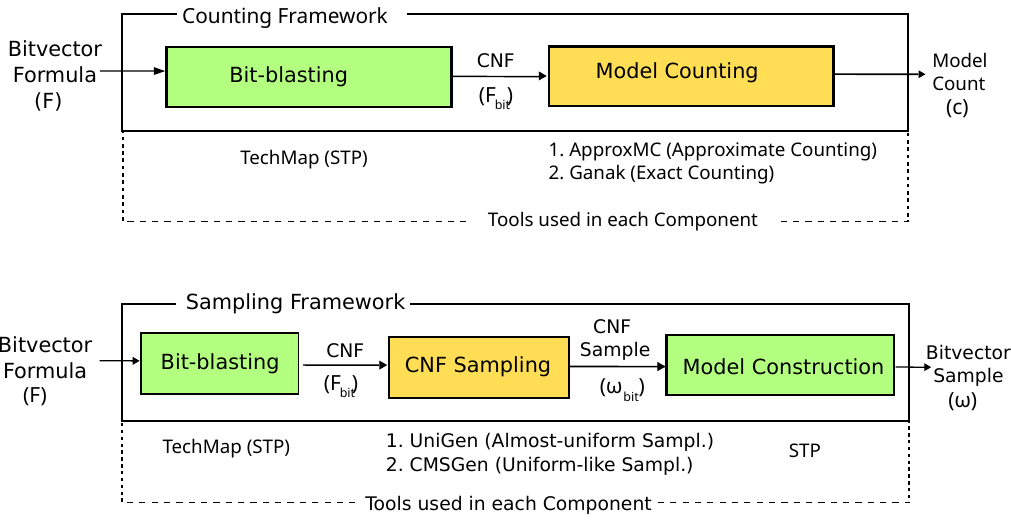}
  \caption{The architecture for counting and sampling in {\tool}.}
  \label{fig:overview}
\end{figure*}

\subsection{Model Counting}

As shown in Figure~\ref{fig:overview}, the model counting part of {\tool} comprises two primary components: (i) a tool for bit-blasting SMT2 formulas into CNFs and (ii) an off-the-shelf model counter.

\paragraph{Phase 1: Bit-blasting}\label{subsec:bit-blasting}
The bit-blasting component takes an input bit-vector formula $F$ and outputs a CNF formula $\fbit$ such that the elements of $\sol(\fbit)$  has one-to-one correspondence with elements of $\sol(F)$. The most commonly used method for converting a bit-vector formula into a propositional Boolean formula involves representing the formula as an And-Inverter Graph (AIG) circuit, {\faig}, which is subsequently converted into CNF, {\fbit}. {\faig} is built with primary input over individual bits of the bitvector formula, {\bits{F}}.
 For conversion to AIG, each bit-vector operator is lowered to a dedicated Boolean gate network over the individual bits. For example, a $w$-bit adder is implemented as a ripple-carry chain of full adders. A $w$-bit multiplier generates $w^2$ partial products  and then sums them using a cascade of adder networks. The AIG $\faig$ is built by composing these per-operator sub-circuits over one another, following the structure of $F$. The result in every case is a directed acyclic graph of and, or, not, xor or ite gates, which are later converted to just and, or gates resulting in the AIG {\faig}.  The AIG is converted to CNF {\fbit} using technology mapping based conversion~\cite{EMS07}. All the auxiliary variables introduced in {\fbit} are defined by bi-implications, therefore, $\sol(\faig) = \solproj{\fbit}{\bits(F)}$, and {\sol(\faig)} has one-one correspondence to elements in {\sol(\fbit)}.

A subset of bits in {\fbit} directly corresponds to the actual bits of the formula $F$, while the rest are auxiliary variables required for encoding constraints. The encoding ensures that the count is not increased by the introduction of the auxiliary variables. These auxiliary variables are not part of the independent support of {\fbit}.  Approximate model counters and samplers achieve better performance when provided with smaller independent support. During bit-blasting, {\csb} identifies and tracks the variables belonging to the independent support versus auxiliary variables, relaying this information to the model counter.

\paragraph{Phase 2: Propositional Model Counting}
The second phase of {\tool} involves a propositional model counter that takes the CNF formula $\fbit$ and returns $c$, the value of $|\sol(\fbit)|$, or an approximation of that. We investigated various approaches to model counting, including both compilation-based counters and hashing-based approximate counters.
In recent years, there has been a proliferation of model counters, and the annual model counting competitions have demonstrated that different model counters have distinct strengths and weaknesses for various problem types. In the context of this research, we examine the applicability of five model counters within the framework of {\tool}: four compilation-based exact counters {\sstd}~\cite{KJ21}, {\gnk}~\cite{SRSM19}, {\dfour}~\cite{LM17}, and {\gpmc}; and a hashing-based approximate counter {\apmc}~\cite{SM19}.
Upon comparing the performances of different CNF counters, we chose the best CNF counters for exact and approximate counting. In the exact model counting mode of {\csb}, we use the latest version of {\gnk} as used in model counting competition 2024~\cite{MCCsubmission}, while in the approximate model counting mode, we use {\apmc}.

\subsection{Projected Model Counting}
The second class of problems {\csb} solves is projected model counting. The architecture is the same as that of non-projected counting, but there are a few key differences.

\paragraph{Phase 1: Bit-blasting}

The bit-blasting component in the projected model counting mode of {\csb} is similar to the non-projected model counting mode. We use {\tm} for bit-blasting. In addition to that, it keeps track of which variables to keep projections on. On the blasted CNF, the variables are marked as projection variables. The CNF counting problem, therefore, represents a projected model counting problem.
Given a bitvector formula $F$, and a projection set $P \subset \vars(F)$, {\csb} detects $P_{\bit}\subset \vars(\fbit)$ from the bitblasted formula {\fbit}, such that $|\solproj{\fbit}{P_\bit}| = |\solprojf{P}|$. The resultant problem becomes a projected CNF counting problem of formula {\fbit}, projected on $P_\bit$.

\paragraph{Phase 2: Propositional Projected Model Counting}
The CNF {\fbit} along with the projection variables $P_\bit$ generated in the previous phase are passed to the projected model counter. In this context, we experimented with four projected model counters {\gnk}~\cite{SRSM19}, {\dfour}~\cite{LM17}, {\gpmc}, and {\apmc}~\cite{SM19}.
Upon comparing the performances of different CNF counters, we chose the best CNF counters for exact and approximate projected CNF counting. Similar to non-projected counting, in the exact model counting mode of {\csb}, we use the latest version of {\gnk}, while in the approximate model counting mode, we use {\apmc}.

\subsection{Uniform Sampling}

As shown in Figure~\ref{fig:overview}, the sampling mode of {\tool} comprises three primary components: (i) a tool for bit-blasting SMT2 formulas into CNFs, (ii) a CNF sampler to sample solutions (almost) uniformly from the solution space, and (iii) constructing the bit-vector model from the sampled CNF solution. In the following part of this section, we briefly describe the task and available tools for each component.

\paragraph{Phase 1: Bit-blasting}\label{subsec:bit-blasting-sampling}
The bit-blasting component generates a propositional formula $\fbit$ from the input bit-vector formula $F$, such that there is a one-to-one correspondence between $\sol(F)$ and $\sol(\fbit)$. The techniques used in this part are the same as those in the bit-blasting component of the model counting mode. At this stage, the underlying SMT solver maintains the mapping between variables of $F$ and $\fbit$, allowing it to map an element of $\sol(\fbit)$ to the corresponding element of $\sol(F)$.

\paragraph{Phase 2: CNF Sampling}
The bit-blasted formula $\fbit$ is passed to a CNF sampler to generate a sample $\sbit$, a randomly selected element from $\sol(\fbit)$. {\unig}~\cite{CFMS+15} uses hashing-based techniques to achieve almost-uniform sampling. On the other hand, {\cmsg}~\cite{GSCM21} incorporates randomization at various stages of an SAT solver's process, effectively making the solver's output mimic that of a uniform sampler. This approach resembles uniform sampling and meets distribution tests~\cite{MPC20}, validating its utility. In practical scenarios, uniform-like samplers often exhibit superior performance, although some applications necessitate strict uniformity guarantees. Depending on the specific sampling requirements, we employ different CNF samplers: for almost-uniform sampling, we utilize {\unig}; for uniform-like sampling, we select {\cmsg}.

\paragraph{Phase 3: Model Construction}
Once the CNF sampler returns a solution $\sbit$, it is passed to the SMT solver, which generates the corresponding bit-vector solution $S$. The SMT solver uses the variable mapping from phase 1 to perform this operation.

\subsection{Preprocessing}
In model counting and sampling, preprocessors attempt to simplify the original problem instance. In practical scenarios, the problem instances are typically rooted in circuits within a specific domain. These circuits are comprised of gates, and the variables correspond to either input or output variables, with the output variables being determined by the input variables. Preprocessing techniques have been developed to effectively handle the \emph{input-output bipartition} property, as discussed in several studies~\cite{LLM16, LLM20, SM22}. In {\tool}, we utilize {\arjun}~\cite{SM22} as a preprocessor, since it consistently outperforms alternative preprocessors~\cite{SM24a}.

\begin{figure}[!t]
  \centering
  \parbox{0.8\textwidth}{
\begin{lstlisting}
(set-logic QF_BV)

(declare-const a (_ BitVec 6))
(declare-const b (_ BitVec 6))
(declare-const c (_ BitVec 6))
(declare-const d (_ BitVec 6))

(declare-projvar a b)
(declare-projvar d)

(assert (bvult a #b001010))
(assert (bvult b #b011110))
(assert (= (bvadd c d) #b001000))

(check-sat)
\end{lstlisting}}
  \caption{Example SMT formula with projection variables}
  \label{fig:smtlib}
\end{figure}

\subsection{Implementation}
A simple approach to build the tool is to use a standalone shell script that interfaces with an SMT solver to generate a CNF file, followed by processing with a model counter or sampler. However, using shell scripts could lead to performance bottlenecks due to the overhead of file read-writes, complicate error handling, and create challenges in deployment and portability by requiring specific environmental setups and external binaries. Therefore, we chose a tightly integrated approach, embedding the counter and samplers directly within the SMT solver as a library. This integration yields a single binary that efficiently produces both model counts and samples. We implement {\tool} using {\stp} as the underlying SMT solver.
The bit-blasting procdure at STP had the following pre-/in-processing steps. We have disabled all of them, as they might have impacted the model-preserving property:
\begin{enumerate}[label=(\roman*)]
\item Bitvector-level simplifications and rewrites: BV-level solving, removal of unconstrained variables, interval analysis, rewrites under \texttt{ite}, equality detection, splitting of \texttt{extract}, share-aware rewrites.
\item Bit-level simplifications: constant-bit propagation, pure-literal elimination, propagation of top-level assertions, flattening of \texttt{and}/\texttt{or}, merging of similar \texttt{or}s, equivalence detection.
\item With these passes disabled, BV$\to$CNF reduces to the two logically count-preserving stages above.
One other possible source that might have broken the model preservation property is variable substitution. However, disabling simplification in {\stp} also disables variable substitution. In {\stp}, substitutions are recorded in a substitution map that is populated  by (i) constant-bit propagation, (ii) pure-literal elimination, and (iii) elimination of unconstrained variables. With these passes disabled, the substitution map remains empty, and variable substitution does not occur.
\end{enumerate}
We integrate {\apmc} to implement the approximate model counter, {\gnk} to integrate the exact model counter, and {\unig} and {\cmsg} to implement the samplers. We also integrate {\arjun} as a preprocessor.  When used alongside {\arjun}, {\tse} and {\tm} exhibit comparable performance; accordingly, we retain {\tm}, the default in {\stp}. The default parameters for the approximate model counting mode are $\varepsilon = 0.8$ and $\delta = 0.2$, adhering to the standard values used in the model counting community. Our implementation is designed to be easily extensible to any (projected) model counter or sampler, requiring only minimal integration effort. However, {\csb} leverages standard APIs for interfacing with counters and samplers; therefore, compatibility requires the availability of an API in the counters or samplers.

\paragraph{Input format}
Since projection is irrelevant in the context of satisfiability, SMT-Lib~\cite{smtlib2} does not incorporate the concept of projection. To address this, we propose an extended format for problem input. Variables can be designated as projection variables using the \texttt{declare-projvar} keyword. Each \texttt{declare-projvar} command can include one or more variables and multiple \texttt{declare-projvar} commands are supported. Projection variables can be declared at any point in the file, provided they are specified after the variable declaration and before the \texttt{check-sat} command. This format is analogous to the one used in propositional model counting~\cite{MCCdataformat}. An example bitvector formula with projection variables is presented in \Cref{fig:smtlib}.

\subsection{Ensuring Correctness}
Given the two-phase structure of {\csb}, correctness of the end-to-end pipeline requires that each phase satisfy its specification.
The expected property from the second phase (CNF counting/sampling) is: given a Boolean formula, count (or sample uniformly) from its satisfying assignments.
The model counters and samplers we use have been extensively validated in the literature~\cite{SM25ganak,CFMS+15,CM19} and in the Model Counting Competition~\cite{mcc2025description};
we therefore treat CNF counting/sampling correctness as a trusted component.
Thus, our primary concern is the correctness of the BV$\to$CNF bit-blasting pipeline.
We establish that this pipeline is model preserving, i.e., the satisfying assignments of the input BV formula $F$ are in one-to-one correspondence with the satisfying assignments of the resulting CNF $\fbit$.
We now review the toolchain to justify this property.
Recall that the BV$\to$CNF pipeline comprises two stages: (i) BV$\to$AIG and (ii) AIG$\to$CNF.
For the two steps, we rely on the following model-preservation properties.

\noindent
\textbf{(1) BV$\to$AIG:}
Recall that bit-blasting is compositional: each operator node in the AST of $F$ is replaced by a dedicated combinational sub-circuit over individual bits, and these sub-circuits are composed along the AST to produce $F_{\mathsf{aig}}$. We establish in two steps that $F$ and $F_{\mathsf{aig}}$ are in one-to-one model correspondence over the primary inputs.
\begin{enumerate}
  \item \textit{{Step 1: Per-operator equivalence.}} For each BV operator, we manually inspect its corresponding AIG circuit. These are standard textbook circuits. Kroening and Strichman~\cite[Ch.~6]{kroening2016decision} establish correctness for many of these circuits by expressing each bit-blasting transformation as a bi-implication between the circuit's output bits and the BV operator's semantics, and proving the resulting iff. We verify that each sub-circuit in our implementation follows this pattern, and is therefore model-preserving. As an empirical safeguard, we further confirm this with 3-bit-width truth-table unit tests, all of which pass.
\item \textit{Step 2: Composition preserves the bijection.} Given Step~1, we lift per-operator equivalence to whole-formula equivalence. Take any satisfying BV assignment $\sigma$ of $F$. Traversing the AST bottom-up, $\sigma$ induces a unique assignment $\sigma'$ to the wires of $F_{\mathsf{aig}}$, and this $\sigma'$ satisfies $F_{\mathsf{aig}}$. Each sub-circuit's output is functionally determined by its inputs, so no spurious models are introduced. Conversely, take any satisfying assignment $\sigma'$ of $F_{\mathsf{aig}}$. Its restriction to the primary inputs is a BV assignment that satisfies $F$, since each sub-circuit's output equals the BV image of its inputs. Hence $\sigma \leftrightarrow \sigma'$ is a bijection between $\mathsf{Sol}(F)$ and $\mathsf{Sol}(\faig)$.
\end{enumerate}

Our testing-based approach to correctness mirrors that of mainstream BV solvers~\cite{ganesh2007decision,boolector}. A fully mechanized argument, in the spirit of recent theorem proving-based efforts for bit-vector reasoning~\cite{SFL+21}, is left to future work.

\noindent
\textbf{(2) AIG$\to$CNF:}
Our toolchain uses technology-mapping-based CNF generation~\cite{EMS07}: it applies function-preserving rewrites to the AIG $\faig$, and then emits a CNF $\fbit$ whose auxiliary variables are constrained by bi-implications.
Consequently, every satisfying input assignment of $\faig$ has a unique extension to a satisfying assignment of $\fbit$; equivalently, $\sol(\faig)$ is in bijection with the projection of $\sol(\fbit)$ onto the input variables.

\paragraph{Emperical assurance.}
We assess the correctness in terms of counting accuracy by comparing counts produced by {\csb} against those generated by {\smap}, the current state-of-the-art bit-vector model counter. As {\smap} is an approximate model counter, direct correctness validation by comparison alone is insufficient. However, if the actual count is $e$, and {\smap} returns a count $s$, we know the actual count $s$ lies within the range $(e/(1+\varepsilon), (1+\varepsilon)e)$ with  probability $(1-\delta)$. Across all benchmark instances where both {\csbe} and {\smap} provided counts, we observed that the results are close enough: $\max\left(\frac{e}{s},\frac{s}{e}\right)-1 < 0.8$,  thereby increasing our confidence in the correctness and reliability of {\csb}.

\section{Experimental Evaluation} \label{sec:results}
The evaluation procedure was conducted using the following setup.
We conducted all our experiments on a high-performance computer cluster, with each node consisting of Intel Xeon Gold 6248 CPUs. We set the memory limit to 16 GB for all configurations and ran each solver instance on a single core. To adhere to the standard timeout used in model counting competitions, we set the timeout for all experiments to 3600 seconds for both counting and sampling.
\footnote{Benchmarks and logfiles available at: \url{https://doi.org/10.5281/zenodo.20637680}}

\paragraph{Baseline} For the problem of model counting, we compare our performance with the prior {\sota} bit-vector model counter, {\smap}. As noted in the related works, there are no available projected model counters or uniform samplers for bit-vectors, so we evaluate the relative performance of both of them based on the total number of benchmarks.

\paragraph{Benchmarks} We collected a substantial set of benchmarks that pertain to the problem of model counting and naturally encode them into bit-vector formulas. The benchmarks were produced using several software tools for various purposes. These include \texsf{CounterSharp}~\cite{TW21}, which is a quantitative software reliability estimation tool; an algorithm designed for testing robust reachability~\cite{GFS21}; \texsf{Delphinium}, which is a cryptographic tool for automating chosen ciphertext attacks~\cite{BZG20}; and \texsf{SMC}, previous work on bit-vector counting~\cite{KM18}. The total number of benchmarks used in our evaluation amounts to {\tot}.
Across the suite, files contain on average 22 bit‑vector variables, 205 constraints and with an overall mean width of 27 bits. The most variables in any one file is 1008, the highest constraint count in a file is 2009, and the single widest bit‑vector declared spans 156 bits.
For projected model counting, we used the same set of benchmarks with the addition of projection variables. In the case of \texsf{CounterSharp} benchmarks, the projection variables were predefined within the benchmarks. For the remaining benchmarks, we randomly selected 30\% of all variables as projection variables.

\mpar
With this setup, in this work, we sought to answer the following questions:
\vspace{.2cm}
\begin{enumerate}[font=\bfseries RQ, leftmargin=1.2cm,label=\arabic*.]
  \item How does {\tool}'s performance compare with the existing state-of-the-art in different problems?
  \item How do the various components of {\tool}, such as model counting, sampling, and bit-blasting, impact its overall performance?
\end{enumerate}

\paragraph{Summary of Results}
In model counting mode, {\tool} solves nearly six times as many instances as {\smap}. Out of a total of {\tot} instances, {\csba} can count {\mxb}, {\csbe} can count {\ecnt}; whereas {\smap} is able to count {\cnts}. In projected counting mode, {\csbe} solves {\ecntp} instances, and {\csba} solves {\acntp} instances.
In sampling mode, {\tool} generated samples from almost all the problem formulas with a median time of 1.17s and 78.4s, respectively, in the uniform-like and almost-uniform sampling mode. The uniform-like sampling mode showed better efficiency in generating samples.

\subsection*{RQ1. Comparison with Prior State-of-the-Art}

As {\csb} solves three completely different problems, we compare the performance in the following three subsections.

\subsubsection*{Problem 1: Model Counting}

\begin{table*}[htb]
  \centering
  \begin{tabular}{@{}lcc@{}}
    \toprule
 Counter & Instances Counted & Average Runtime (s) \\ \midrule
    \smap{} & {\cnts}               & 5973.6              \\
 {\csbe} & {\ecnt} & {2795.9} \\
 {\csba} & \textbf{{\mxb}}   & \textbf{368.3}      \\\bottomrule
  \end{tabular}
  \caption{Performance comparison on {\tot} non-projected instances.}
  \label{tab:countnonproj}
  \end{table*}

We experimented with configurations for {\tool} and compared its performance with {\smap}, the current state-of-the-art approximate bit-vector model counter. \Cref{tab:countnonproj} shows the improvement in performance demonstrated by {\tool}. In approximate mode, {\tool} counted {\mxb} instances out of {\tot}, which is six times the number of instances counted by {\smap}, which could count {\cnts} instances. Even in exact mode, {\csb} performs better than {\smap}, which is an approximate counter.
The average Runtime\footnote{We use the PAR-2 score as a proxy for average runtime. PAR-2 score (penalized average runtime) assigns a runtime of two times the time limit for each benchmark not solved by a counter.} of {\csb} is also great.

\begin{figure}[htb]
  \centering
  \resizebox{0.6\linewidth}{!}{\input{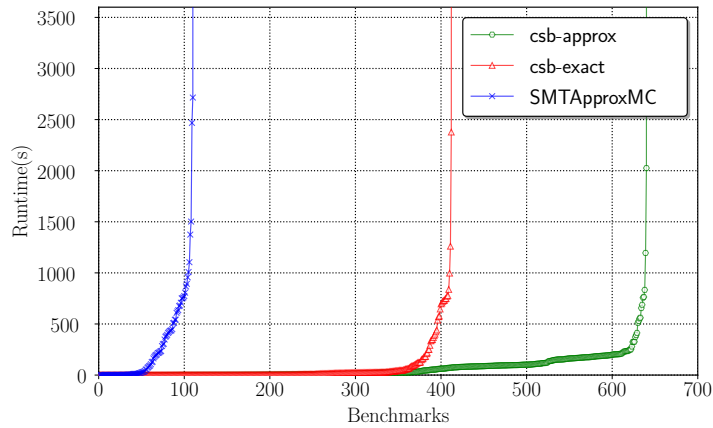}}
  \caption{Performance of {\csb} with {\smap} on non-projected counting. }
  \label{fig:cactusnonproj}
  \end{figure}

In the cactus plot in \Cref{fig:cactusnonproj}, we compare the performance. The $x$-axis represents the number of instances, while the $y$-axis shows the time taken. A point $(i, j)$ in the plot indicates that a counter counted $j$ benchmarks out of the total benchmarks in a set in less than or equal to $i$ seconds. The cactus plot reveals that while {\smap} struggles to solve more than {\cnts} instance, {\csb} solves more than 600 instances, each with less than 500s.

\subsubsection*{Problem 2: Projected Model Counting}

\begin{figure}[htb]
  \centering
  \resizebox{0.75\linewidth}{!}{\input{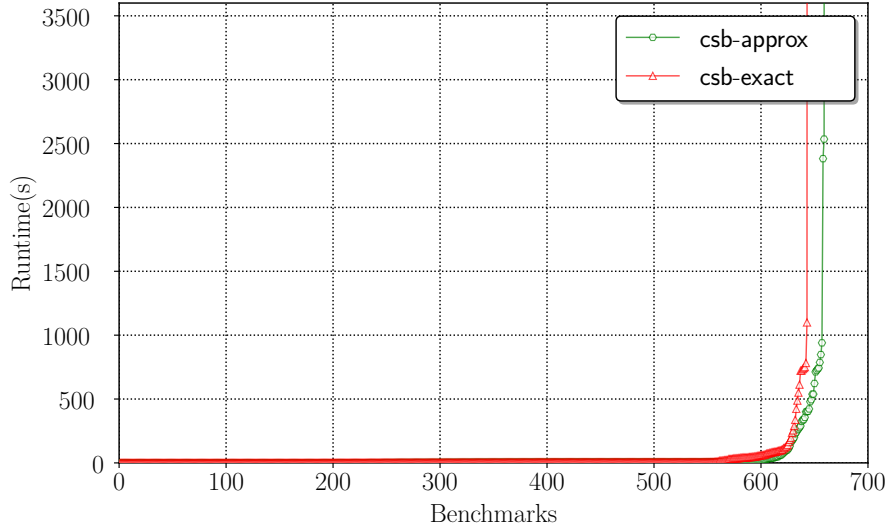}}
  \caption{Performance comparison of two modes of {\csb} on projected counting. }
  \label{fig:cactusproj}
\end{figure}

As there is no existing uniform bit-vector sampling tool, we do not have a baseline for comparing {\tool} against. We count the number of instances solved by the approximate and exact mode of {\csb} out of {\tot} instances. As \Cref{tab:countproj} indicates, in approximate mode {\csb} solves {\acntp} instances - almost all of the benchmark set consisting of {\tot} instances. The performance in exact mode is also quite close, where {\csb} solves {\ecntp} instances. Moreover, the cactus plot in \Cref{fig:cactusproj} indicates that in both modes, {\csb} solves 607 instances in a minute.

\begin{table*}[htb]
  \centering
  \begin{tabular}{@{}lcc@{}}
    \toprule
 Counter & Instances Counted & Average Runtime (s) \\ \midrule
 {\csbe} & {\ecntp}          & 301.4               \\
 {\csba} & \textbf{\acntp}   & \textbf{140.2}      \\\bottomrule
  \end{tabular}
  \caption{Performance comparison on {\tot} projected instances.}
  \label{tab:countproj}
\end{table*}

\subsubsection*{Problem 3: Uniform Sampling}

As there is no existing uniform bit-vector sampling tool, we do not have a baseline for comparing {\tool} against. We tested {\tool} on our benchmarks in both sampling modes by asking it to generate 500 samples for each input formula. In \Cref{tab:performance}, we compare the performance in terms of the number of instances solved and average runtime. Out of {\tot} instances, in almost-uniform sampling mode {\tool} generated samples for 641 instances, while in uniform-like sampling mode, it solved 660 instances.

\begin{table*}[htb]
  \begin{tabular}{lccc}
    \toprule
 Sampling Mode in {\tool} & Instances Sampled & Avg. Time (s) & Median Time (s) \\ \midrule
 Almost-uniform sampling   & 641               & 283.7         & 78.4            \\
 Uniform-like sampling     & 660               & 6.6           & 1.17            \\ \bottomrule
  \end{tabular}
  \caption{Performance comparison on sampling on {\tot} instances.}
  \label{tab:performance}
\end{table*}

\subsection*{RQ2. Impact of Different Components}
As shown in \Cref{sec:framework}, there are multiple components for {\tool}, and there are multiple tools/algorithms available for each component. To determine which tools to use for each component, we performed experiments. Below, we summarize the impact of each component.

\paragraph{Impact of CNF Counters in non-projected counting} To find the best counters to use in {\csb}, we selected solvers from the 2024 Model Counting Competition~\cite{MCCsubmission} for comparison. By combining each model counter with its best possible preprocessing settings, we observed that {\apmc} delivered the best performance, solving {\mxb} out of {\tot} benchmarks. The second-best performance was achieved by {\gnk}, which solved {\ecnt} benchmarks. The other exact counters — {\sstd}, {\dfour}, and {\gpmc} — performed similarly to {\gnk}, solving 414, 359, and 342 instances, respectively. As shown in the cactus plot in \Cref{fig:cactusnonproj}, {\apmc} outperforms all other counters significantly, while the exact counters exhibit comparable performance.

\begin{figure}[htb]
  \centering
  \resizebox{0.75\linewidth}{!}{\input{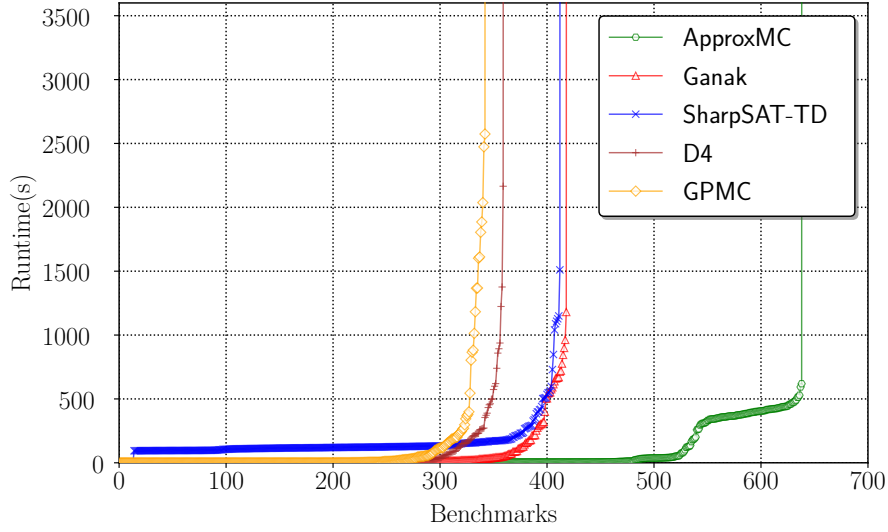}}
  \caption{Performance comparison of model counters on non-projected instances.}
  \label{fig:cactus-counters-nonproj}
\end{figure}

\begin{figure}[htb]
  \centering
  \resizebox{0.75\linewidth}{!}{\input{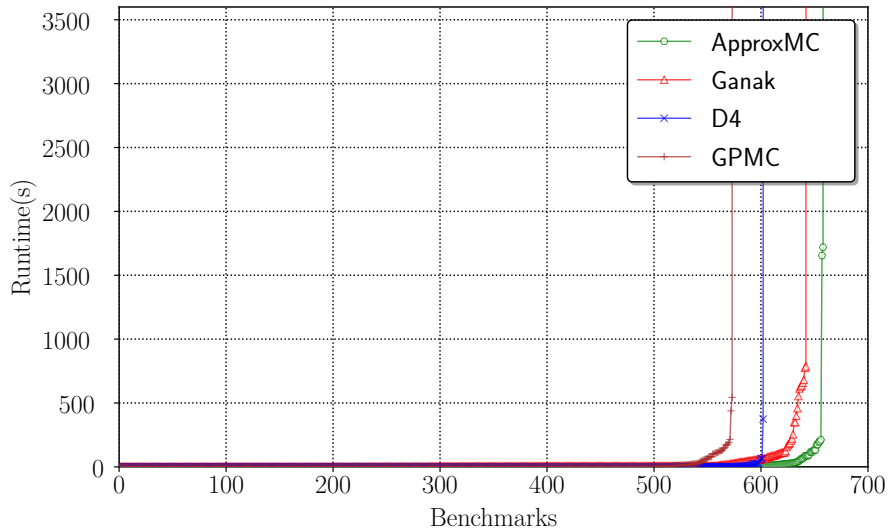}}
  \caption{Performance comparison of model counters on projected instances.}
  \label{fig:cactus-counters-proj}
\end{figure}

\paragraph{Impact of CNF Counters in projected counting} Similar to the non-projected case, we evaluate projected CNF counters on the projected CNF counting instances generated by {\csb}. The counter {\sstd} was excluded from this evaluation as it does not support projected model counting. {\apmc} achieved the best performance, solving {\acntp} instances. Among the exact counters, {\gnk} performed the best, solving {\ecntp} instances. The other counters showed similar performance: {\dfour} solved 602 instances, and {\gpmc} solved 573 instances. The cactus plot in \Cref{fig:cactus-counters-proj} shows that, unlike the non-projected case, the performance gap between {\apmc} and the exact counters is minimal, with all counters performing comparably.

\paragraph{Impact of CNF Samplers} Both CNF samplers demonstrate strong performance, solving over 94\% of instances. The average runtime is significantly improved when {\cmsg} is used as the sampler. In \Cref{tab:performance}, we compare the performance metrics. Using {\cmsg} as the sampler, {\tool} solves 21 more instances and reduces the average runtime to 3\% of that with {\unig}.

\paragraph{Impact of bitblasting method}
There are multiple techniques available for converting an AIG to CNF, with the two most commonly used being: (i) {\tse}: the standard Tseitin encoding~\cite{Tse83} and (ii) {\tm}: technology mapping-based logic synthesis~\cite{EMS07}. The {\tse} method uses auxiliary variables for sub-circuits of AIG, whereas the {\tm} method performs various optimizations on the circuit to produce a minimized CNF. From the perspective of satisfiability, studies have revealed that the performance of the encoding scheme is reliant on the benchmark set being investigated, with {\tse} exhibiting superior results in some benchmark sets and {\tm} being more effective in others~\cite{JLS09}. In the context of  {\tool}, {\tse} and {\tm} showed similar performance, and we use {\tm} for bit-blasting, which is default in {\stp}.
The performance of {\tool} remains relatively unchanged when the bit-blasting technique is switched between {\tse} and {\tm} in {\csb}. However, if $\mathsf{Arjun}$ is not employed as a preprocessing technique, using {\tse} as a bit-blasting technique instead of {\tm} can improve the performance of {\tool}. This improvement is observed across all CNF-counters.

\paragraph{Impact of Preprocessing} The preprocessors had a minimal positive impact on the performance of exact model counting tools, and in a few cases, they even showed a slight negative effect. For the approximate model counting tool {\apmc}, {\arjun} demonstrated a positive impact, enabling the solving of 643 instances, which is 473 more than without preprocessing.

\section{Conclusion} \label{sec:concl}

This work introduced {\tool}, an extension of the SMT solver {\stp} that utilizes CNF counters and samplers to efficiently address counting and sampling challenges over bit-vector formulas efficiently, demonstrating strong performance on a large application benchmark set. {\tool} supports exact and approximate projected and non-projected model counting, as well as almost-uniform and uniform-like sampling, marking the first solution for exact counting, projected counting, and uniform sampling on bit-vector formulas. We hope that the availability of an efficient tool will inspire researchers from diverse domains to solve their problems using bit-vector counting and sampling, fostering further exploration and innovation in this area.

\paragraph*{Acknowledgements.}
We are thankful to anonymous reviewers for their constructive feedback. This work was supported in part by the Natural Sciences and Engineering Research Council of Canada (NSERC) [RGPIN-2024-05956]. The work was done when Arijit Shaw was a visiting graduate student at the University of Toronto. Computations were performed on the Niagara supercomputer at the SciNet HPC Consortium. SciNet is funded by Innovation, Science and Economic Development Canada; the Digital Research Alliance of Canada; the Ontario Research Fund: Research Excellence; and the University of Toronto.

\clearpage

\bibliographystyle{alpha}
\bibliography{bib}

\end{document}